\UseRawInputEncoding
\documentclass[aps,prb,amsmath,amssymb,reprint]{revtex4-2}
\usepackage{graphicx}% Include figure files
\usepackage{dcolumn}% Align table columns on decimal point
\usepackage{bm}% bold math
\usepackage[colorlinks,bookmarksopen,bookmarksnumbered,citecolor=blue, linkcolor=blue, urlcolor=blue]{hyperref}

\begin{document}

%\preprint{}

\title{The skyrmion bags in an anisotropy gradient}


\author{Zhaozhuo Zeng$^{1}$}
\author{Nasir Mehmood$^{1}$}
\author{Yunxu Ma$^{1}$}
\author{Jianing Wang$^{1}$}
\author{Jianbo Wang$^{1,2}$}
\author{Qingfang Liu$^{1,}$}
\email{liuqf@lzu.edu.cn}
\affiliation{$^{1)}$Key Laboratory for Magnetism and Magnetic Materials of the Ministry of Education, Lanzhou University, Lanzhou 730000, People’s Republic of China.\\$^{2)}$Key Laboratory for Special Function Materials and Structural Design of the Ministry of Education, Lanzhou University, Lanzhou 730000, People’s Republic of China.}

%\thanks{}


\date{\today}

\begin{abstract}
 Skyrmion bags as spin textures with arbitrary topological charge are expected to be the carriers in racetrack memory. Here, we theoretically and numerically investigated the dynamics of skyrmion bags in an anisotropy gradient. It is found that, without the boundary potential, the dynamics of skyrmion bags are dependent on the spin textures, and the velocity of skyrmionium with $Q = 0$ is faster than other skyrmion bags. However, when the skyrmion bags move along the boundary, the velocities of all skyrmion bags with different $Q$ are same. This can be attributed to the same value of $u/\eta_{xx}$, where the $u$ and $\eta_{xx}$ are the terms related to the magnetization distribution of skyrmion bag. In addition, we theoretically derived the dynamics of skyrmion bags in the two cases using the Thiele approach and discussed the scope of Thiele equation. Within a certain range, the simulation results are in good agreement with the analytically calculated results. Our findings provide an alternative way to manipulate the racetrack memory based on the skyrmion bags.
 \end{abstract}


\maketitle

%\section{}
%\subsection{}
%\subsubsection{}
\section{Introduction}

 Since its discovery, the skyrmion\cite{R1,R2}, being a topological quasiparticle, is suggested to be used in spintronic devices, such as racetrack memories\cite{R3,R4,R5}, logic devices\cite{R6,R7,R8}, and artificial neuron devices\cite{R9,R10,R11}, etc. How to manipulate this kind of structure is a very important problem that scientific community is concerned about. One of the most common method is to use the spin-polarized current to drive their movement in the nanodevices\cite{R12,R13,R14,R15}. However, this kind of method has the drawbacks of generating Joule heat, large energy consumption, low energy conversion efficiency and inability to be driven in magnetic insulators. To overcome these shortcomings, alternative methods have been suggested, like using spin wave\cite{R16,R17}, magnetic field gradient\cite{R18,R19}, temperature gradient\cite{R20,R21}, or microwave field\cite{R22,R23} instead of spin-polarized current to manipulate these topological structures. Besides that, another method that has been proved to be efficient in controlling the spin textures is based on the voltage-controlled magnetic anisotropy (VCMA) effect\cite{R24,R25}. Based on this strategy, the magnetic anisotropy gradient has been widely used in the manipulation of spin textures\cite{R26,R27,R28,R29,R30,R31,R32,R33,R34,R35}. Skyrmion bags, as spin textures with arbitrary topological charge\cite{R36,R37,R38}, are expected to become the information carriers for multiple-data and high-density racetrack memory\cite{R39,R40,R41} and interconnect device\cite{RN1,RN2}. In order to present an alternative way to manipulate the skyrmion bags with higher efficiency and larger controllability, we investigated the dynamics of skyrmion bags in an anisotropy gradient.

\begin{figure}
\includegraphics[width=1\columnwidth]{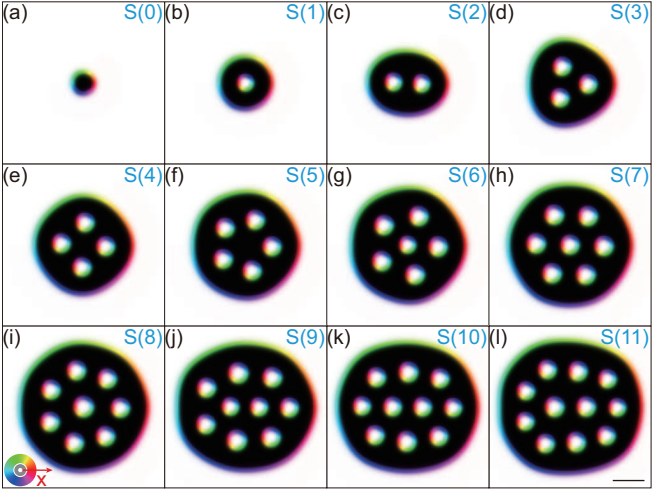}%
\caption{\label{Fig:1}(a)-(l) Schematics of magnetization distribution from S(0) to S(11). The white and black represent the vertical components of the magnetization pointing up and down, respectively. The color wheel in (i) indicates the in-plane component and the scale bar in (l) represents 50 nm. Material parameters used are: the saturation magnetization $M_s$ is 580 kA/m, the exchange constant $A$ is 15 pJ/m, the anisotropy constant $K$ is 0.8 $\mathrm{MJ/m^3}$, and the interface DMI constant $D$ is 3.5 $\mathrm{mJ/m^2}$, respectively.}
\end{figure}

 It is noticed that for skyrmion bags with the same topological number, the different geometries and symmetries lead to different dynamics\cite{R43}. In order to reduce the influence to the dynamics in our investigation, the skyrmion bags in the present study have similar magnetic textures, where a skyrmion cluster is surrounded by the large skyrmion outer boundary, as shown in Fig.~\ref{Fig:1}. We use the number of small skyrmions ($N$) in skyrmion cluster to mark the different skyrmion bags, i.e. S($N$). When we consider the topological charge ($Q$) of a skyrmion bag, with $Q$ defined as:
 \begin{equation}
 	Q = \frac{1}{4\pi}\int{\bm{m}\cdot{\frac{\partial\bm{m}}{\partial x}}\times{\frac{\partial\bm{m}}{\partial y}}}dxdy, \label{eq:1}
 \end{equation}
a skyrmion bag is marked as S($Q$+1). For example, the single skyrmion with $Q = -1$ is marked as S(0) and the skyrmionium with $Q = 0$ is marked as S(1). Additionally, with the $N$ increasing, the size of a skyrmion bag increases due to the repulsive interaction.

\section{Model and Method}

In our study, we used the micromagnetic simulation software Mumax3 to perform the dynamics of skyrmion bags in an anisotropy gradient with Landau-Lifshitz-Gilbert equation\cite{R44}:
\begin{equation}
	\frac{d\bm{m}}{dt} = -\frac{\gamma}{1+\alpha^{2}}\bm{m}\times\bm{H}_{eff}-\frac{\alpha\gamma}{1+\alpha^{2}}\bm{m}\times(\bm{m}\times \bm{H}_{eff}), \label{eq:2}
\end{equation}
where $\gamma$ is the gyromagnetic radio, $\alpha$ is the Gilbert damping constant and $\bm{H}_{eff}$ is the effective field. In an anisotropy gradient, the $\bm{H}_{eff}$ can be expressed as\cite{R26,R28}:
\begin{eqnarray}
	\bm{H}_{eff} =&& \frac{2}{\mu_0 M_s}[A\nabla^2\bm{m}+(K_m-x\cdot\vert dK/dx \vert)m_z\bm{e}_z\nonumber\\
	&&+D(\nabla \bm{m}_z-(\nabla \bm{m})\bm{e}_z)]+\bm{H}_d, \label{eq:3}
\end{eqnarray}
where $M_s$ is the saturation magnetization, $A$ is the Heisenberg exchange constant, $K_m$ is the magnetic anisotropy constant at the center of magnetic film, $\vert dK/dx \vert$ is the magnitude of anisotropy gradient, $D$ is the interfacial DMI constant, and $\bm{H}_d$ is the demagnetization field. In our study, we investigated the skyrmion bags in Co/Pt film\cite{R28,R39,R42}, where $M_s = 580\ \mathrm{kA/m}$, $A = 15\ \mathrm{pJ/m}$, and $D = 3.5\ \mathrm{mJ/m^2}$.

\begin{figure}
	\includegraphics[width=1\columnwidth]{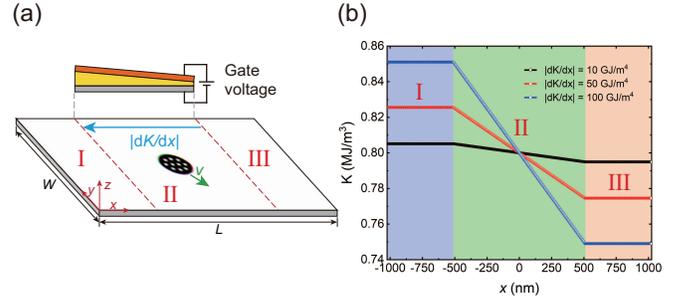}%
	\caption{\label{Fig:2}(a) Schematic of S(10) induced by an anisotropy gradient along $-x$ direction. The $L$ and $W$ are the length and width of magnetic film. Region \uppercase\expandafter{\romannumeral1} and \uppercase\expandafter{\romannumeral3} are buffer zones with a fixed anisotropy constant and Region \uppercase\expandafter{\romannumeral2} is the gradient-driven region with an anisotropy gradient. The sandwich structure is the side view of Region \uppercase\expandafter{\romannumeral2}, where the grey, gold, and orange represent the magnetic layer, insulator layer, and electrode layer, respectively. (b) The magnitude of magnetic anisotropy $K$ as a function of \textit{x}-coordinate with $L$ = 2048 nm.}
\end{figure}

As shown in Fig.~\ref{Fig:2}(a), we considered a magnetic film with length $L$, width $W$ and the thickness of 1 nm. The mesh is $2\times2 \times 1 \ \mathrm{nm^3}$. In order to reduce the influence of the boundary potential from both sides of the film on the dynamics of skyrmion bags, Region \uppercase\expandafter{\romannumeral1} and \uppercase\expandafter{\romannumeral3} as buffer zone have a fixed anisotropy constant with different values. The area between two red dotted lines is the gradient-driven region (Region \uppercase\expandafter{\romannumeral2}) with an anisotropy gradient, which is induced by VCMA and the insulator layer with thinkness gradient, as shown in Fig.~\ref{Fig:2}(a). The width of Region \uppercase\expandafter{\romannumeral1} and \uppercase\expandafter{\romannumeral3} is $L/4$, while that of Region \uppercase\expandafter{\romannumeral2} is $L/2$. All of the results are obtained in Region \uppercase\expandafter{\romannumeral2}. It is worth noting that in Region \uppercase\expandafter{\romannumeral2} the magnitude of magnetic anisotropy is defined as $K = K_m-x\cdot\vert dK/dx \vert$, where $K_m$ is set to be $0.8\, \mathrm{MJ/m^3}$. Figure~\ref{Fig:2}(b) is the magnitude of magnetic anisotropy $K$ as a function of \textit{x}-coordinate. When $\vert dK/dx \vert  = 10\ \mathrm{GJ/m^4}$ and $L = 2048\ \mathrm{nm}$, the anisotropy constants of Region \uppercase\expandafter{\romannumeral1} and \uppercase\expandafter{\romannumeral3} are $0.80512\ \mathrm{MJ/m^3}$ and $0.79488\ \mathrm{MJ/m^3}$, respectively. While the anisotropy constants of Region \uppercase\expandafter{\romannumeral1} and \uppercase\expandafter{\romannumeral3} are $0.8512\ \mathrm{MJ/m^3}$ and $0.7488\ \mathrm{MJ/m^3}$ for $\vert dK/dx \vert  = 100\ \mathrm{GJ/m^4}$, respectively.

Subsequently, we describe the dynamics of skyrmion bags in an anisotropy gradient with and without the boundary potential using the Thiele approach\cite{R45}. From Eq.~(\ref{eq:2}) and (\ref{eq:3}), we can derive the Thiele equation for skyrmion bags in an anisotropy gradient\cite{R26,R27,R28} as:
\begin{equation}
	\bm{G \times v}+\alpha\eta\bm{v}+\frac{\gamma u}{\mu_0 M_s}\vert dK/dx \vert \bm{e}_x = \nabla U(y), \label{eq:4}
\end{equation}
where $G = -4\pi Q \bm{e}_z$ is gyromagnetic coupling vector, \begin{math}
\eta = \left( 
\begin{smallmatrix}
\eta_{xx} & \eta_{xy} \\ \eta_{yx} & \eta_{yy}
\end{smallmatrix}
\right)
\end{math} is the dissipation tensor, the last term on the left-hand side of Eq.~\ref{eq:4} is the driving force along $x$ direction generated by the magnetic anisotropy gradient, and  $\nabla U(y)$ is the force generated by the boundary potential. It is worth noting that the components of $\eta$ and the term $u$ are related to the magnetization distribution of skyrmion bag and can be defined as:
\begin{equation}
	\eta_{ij} = \int\partial_i \bm{m}(\bm{r})\cdot\partial_j \bm{m}(\bm{r})d^2r, \label{eq:5}
\end{equation}
\begin{equation}
	u = \int(1-m_z^2)dxdy. \label{eq:6}
\end{equation}
When the boundary potential is absent, i.e. $\nabla U(y) = 0$, we obtain $v_x$ and $v_y$ of skyrmion bag as:
\begin{equation}
\left\lbrace
\begin{aligned}
v_x &= \frac{\alpha\eta_{yy}}{G^2+\alpha^2\eta_{xy}\eta_{yx}+\alpha\eta_{xx}\alpha\eta_{yy}}\frac{\gamma u}{\mu_0 M_s}\vert dK/dx \vert\\
v_y &= \frac{G-\alpha\eta_{yx}}{G^2+\alpha^2\eta_{xy}\eta_{yx}+\alpha\eta_{xx}\alpha\eta_{yy}}\frac{\gamma u}{\mu_0 M_s}\vert dK/dx \vert
\end{aligned}.
\right. \label{eq:7}
\end{equation}
We can also obtain the velocity ($v$) and skyrmion Hall angle ($\theta_{Sky}$) of skyrmion bag as:
\begin{equation}
\left\lbrace
\begin{aligned}
&v = \sqrt{v_x^2+v_y^2}\\
&\theta_{Sky} = \arctan\left(\frac{v_y}{v_x}\right) = \arctan\left(\frac{G-\alpha\eta_{yx}}{\alpha\eta_{yy}}\right) 
\end{aligned}.
\right. \label{eq:8}
\end{equation}
If the boundary potential is nonzero and skyrmion bag moves along the boundary, due to the balance between the magnus force and the force generated by the boundary, the velocity component along y direction ($v_y$) can be assumed to be zero\cite{R39}. Hence, we can obtain the velocity of skyrmion bag moving along the boundary as:
\begin{equation}
	v = v_x = \frac{\gamma u}{\alpha\eta_{xx}\mu_0 M_s}\vert dK/dx \vert. \label{eq:9}
\end{equation}

\section{Results and discussion}

\subsection{The skyrmion bags in different magnetic anisotropy constants}

\begin{figure}
	\includegraphics[width=1\columnwidth]{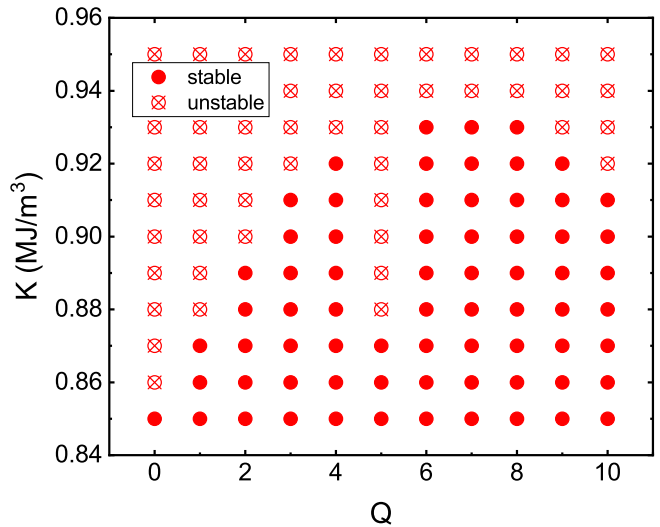}%
	\caption{\label{Fig:3} Stability of skyrmion bags with topological number $Q$ from 0 to 10 under different magnetic anisotropy constants $K$. The red solid circle represents that the skyrmion bag is stable, and the red open circle with a cross represents that the skyrmion bag is unstable.}
\end{figure}

Due to the inner interactions in skyrmion bags, the maximum magnetic anisotropy constant allowed to exist is the key to the dynamics of skyrmion bags driven by an anisotropy gradient\cite{R38}. Hence, we investigated the maximum magnetic anisotropy constant allowed for the stable existence of skyrmion bags with topological charge $Q$ ranging from 0 to 10, as shown in Fig.~\ref{Fig:3}. Among them, the red solid circle represents that the skyrmion bag is stable and the topological charge does not change over time, while the red open circle with a cross represents that the skyrmion bag is unstable and the topological charge changes over time. It is found that the maximum magnetic anisotropy constant of skyrmionium is the smallest as compared with other skyrmion bags, which is attributed to the small size of skyrmionium. With the increasing $Q$, the maximum magnetic anisotropy constant increases due to the inner interactions. However, the maximum magnetic anisotropy constant drops suddenly with $Q$ = 5. Due to the interactions between the central skyrmion and other small skyrmions for S(6), it is easily annihilated. Subsequently, with the increasing $Q$, the coupling between small skyrmions on the same circle gradually increases\cite{R42}, resulting in an increase in the maximum magnetic anisotropy constant. For S(9)-S(11), due to the reduced rotational symmetry, the maximum magnetic anisotropy constant gradually decreases with the increase of $Q$.

\subsection{Dynamics of skyrmion bags in an anisotropy gradient without boundary potential}

\begin{figure}
	\includegraphics[width=1\columnwidth]{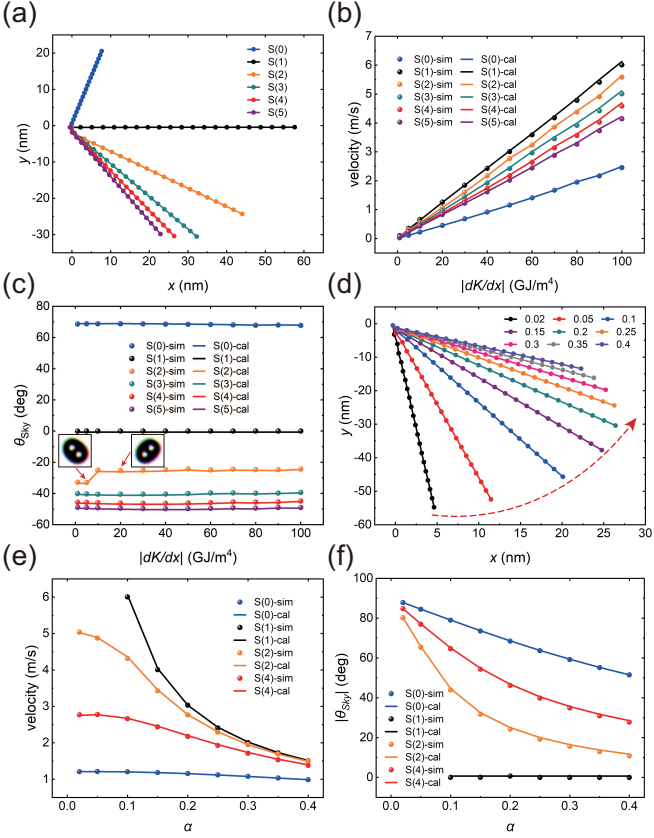}%
	\caption{\label{Fig:4} (a) The topological trajectory of S(0)-S(5) within 20 ns when $\alpha = 0.2$ and $\vert dK/dx \vert = 50\ \mathrm{GJ/m^4}$. The time interval between two dots is 1 ns. (b) The velocity of S(0)-S(5) as a function of $\vert dK/dx \vert$. The dots are the simulated data, and the line is the data calculated from Eq.~(\ref{eq:8}). (c) The skyrmion Hall angle $\theta_{Sky}$ of S(0)-S(5) as a function of $\vert dK/dx \vert$. The illustrations show the schematics of magnetization distribution of S(2) when moving with $\vert dK/dx \vert = 5\ \mathrm{GJ/m^4}$ and $\vert dK/dx \vert = 20\ \mathrm{GJ/m^4}$, respectively. (d) The topological trajectory of S(4) within 20 ns under different Gilbert damping constants when $\vert dK/dx \vert = 50\ \mathrm{GJ/m^4}$. The time interval between two dots is 1 ns. (e) The velocity of skyrmion bags as a function of $\alpha$. (f) The deflection angle $\vert \theta_{Sky} \vert$ of skyrmion bags as a function of $\alpha$.}
\end{figure}

In order to eliminate the influence of boundary potential, the periodic boundary condition is added in $y$ direction. The length ($L$) along \textit{x}-direction is set to be 2048 nm and the width ($W$) along \textit{y}-direction is set to be 1024 nm. The S(0)-S(5) move along straight lines with different skyrmion Hall angles ($\theta_{Sky}$) when $\alpha = 0.2$ and $\vert dK/dx \vert = 50\ \mathrm{GJ/m^4}$, as shown in Fig.~\ref{Fig:4}(a). Unsurprisingly, S(1) moves along a horizontal straight line. For S(2) to S(5), the deflection angle $\vert \theta_{Sky} \vert$ increases and the velocity decreases gradually with increasing $Q$. In the case of S(0), a single skyrmion with $Q = -1$, it has the largest skyrmion Hall angle and the smallest velocity. It is worth noting that because the $Q$ of S(0) and S(2) are opposite numbers, the signs of $\theta_{Sky}$ are also opposite. 

From Eq.~(\ref{eq:7}) and (\ref{eq:8}), it is found that the velocity of skyrmion bag is proportional to $\vert dK/dx \vert$, while the $\theta_{Sky}$ is independent of $\vert dK/dx \vert$. Hence, we first study the velocity of S(0)-S(5) as a function of $\vert dK/dx \vert$ and the result is shown in Fig.~\ref{Fig:4}(b). The dots are the simulated data, and the line is the data calculated from Eq.~(\ref{eq:8}). It is found that the velocity of skyrmion bags increases linearly with the $\vert dK/dx \vert$ increasing. Moreover, at the same $\vert dK/dx \vert$, the velocity of S(1) is the largest, while that of S(0) is the smallest, and the velocity of other skyrmion bags is in between these two cases, and gradually becomes smaller as the $Q$ increases. When $\vert dK/dx \vert = 50\ \mathrm{GJ/m^4}$, the velocities of S(0)-S(5) are 1.15, 3.03, 2.78, 2.45, 2.20, and 2.04 m/s, respectively. Next, we considered the $\theta_{Sky}$ of S(0)-S(5) as a function of $\vert dK/dx \vert$, as shown in Fig.~\ref{Fig:4}(c). It is found that the $\theta_{Sky}$ of skyrmion bags is almost unchanged with the $\vert dK/dx \vert$ increasing. The $\theta_{Sky}$ of S(0)-S(5) is approximately  $68^{\circ}$, $0^{\circ}$, $-25^{\circ}$, $-40^{\circ}$, $-46^{\circ}$, and $-50^{\circ}$, respectively. Interestingly, the $\theta_{Sky}$ of S(2) has a jump with $\vert dK/dx \vert = 10\ \mathrm{GJ/m^4}$. The illustrations in Fig.~\ref{Fig:4}(c) show the schematics of magnetization distribution when $\vert dK/dx \vert = 5\ \mathrm{GJ/m^4}$ and $\vert dK/dx \vert = 20\ \mathrm{GJ/m^4}$, respectively. Due to different $\eta_{yx}$ of different magnetization distributions, their $\theta_{Sky}$ are different from Eq.~\ref{eq:8}\cite{R43}.

Moreover, we studied the dynamics of S(0), S(1), S(2), and S(4) under different $\alpha$ with $\vert dK/dx \vert = 50\ \mathrm{GJ/m^4}$. Figure~\ref{Fig:4}(d) is the topological trajectory of S(4) under different $\alpha$. It is found that with the increase in $\alpha$, both the velocity and $\vert \theta_{Sky} \vert$ decrease, as shown by the red dotted line with arrow in Fig.~\ref{Fig:4}(d). Furthermore, we investigated the velocity of skyrmion bags as a function of $\alpha$, as shown in Fig.~\ref{Fig:4}(e). It is found that the velocity of skyrmion bags decreases with the increase of $\alpha$ and the velocity of skyrmionium is more dependent on $\alpha$ when $\alpha < 0.4$. When $\alpha$ = 0.1, the velocities of S(0), S(1), S(2), and S(4) are 1.20, 6.00, 4.32, and 2.66 m/s, respectively. When $\alpha$ = 0.4, the velocities of S(0), S(1), S(2), and S(4) are 0.99, 1.50, 1.49, and 1.38 m/s, respectively. Figure~\ref{Fig:4}(f) shows the deflection angle $\vert \theta_{Sky} \vert$ of skyrmion bags as a function of $\alpha$. With the increase in $\alpha$, the $\vert \theta_{Sky} \vert$ decreases. The differences of deflection angle of S(1), S(2), and S(4) between $\alpha = 0.02$ and $\alpha = 0.4$ are $36.31^{\circ}$, $69.30^{\circ}$, and $56.95^{\circ}$, respectively. 

Next, we analytically studied the velocity and $\theta_{Sky}$ of S(0)-S(11) when $\alpha = 0.2$ and $\vert dK/dx \vert = 50\ \mathrm{GJ/m^4}$, and the result is shown in Table \ref{tab:1}. In order to evaluate the accuracy of Eq.~(\ref{eq:8}), the relative error parameter of velocity $\varepsilon_v$ is defined as:
\begin{equation}
	\varepsilon_v = \vert \frac{v_{cal}-v_{sim}}{v_{sim}} \vert \times 100\%, \label{eq:10}
\end{equation}
where $v_{cal}$ and $v_{sim}$ are the calculated and simulated velocity, respectively. It can be found that all of $\varepsilon_v$ is less than 1.5 \%, which indicates that the simulation results are in good agreement with the analytically calculated results. Additionally, with the increasing $Q$, the velocity and $\theta_{Sky}$ of S(2) to S(11) decrease, and the rate of change gradually decreases. It indicates that although the skyrmion bags have arbitrary topological charge, there should be corresponding limits on the velocity and $\theta_{Sky}$\cite{R40,R41}.

\begin{table}[b]%The best place to locate the table environment is directly after its first reference in text
	\caption{\label{tab:1}%
		The topological charge $Q$, simulated velocity $v_{sim}$, calculated velocity $v_{cal}$, relative error parameter of velocity $\varepsilon_v$, simulated skyrmion Hall angle $\theta_{Sky-sim}$, and calculated skyrmion Hall angle $\theta_{Sky-cal}$ for each bag with $\alpha = 0.2$ and $\vert dK/dx \vert = 50\ \mathrm{GJ/m^4}$.
	}
	\begin{ruledtabular}
		\begin{tabular}{ccccccc}
			\textrm{Bag}&
			\textit{Q}&
			\multicolumn{1}{c}{$\begin{aligned}&\ v_{sim}\\&\mathrm{(m/s)}\end{aligned}$}&
			\multicolumn{1}{c}{$\begin{aligned}&\ v_{cal}\\&\mathrm{(m/s)}\end{aligned}$}&
			\multicolumn{1}{c}{$\varepsilon_v\, \mathrm{(\%)}$}&
			\multicolumn{1}{c}{$\begin{aligned}&\theta_{Sky-sim}\\&\quad\mathrm{(deg)}\end{aligned}$}&
			\multicolumn{1}{c}{$\begin{aligned}&\theta_{Sky-cal}\\&\quad\mathrm{(deg)}\end{aligned}$}\\
			\hline
			\textrm{S(0)} & -1 & 1.16 & 1.15 & 0.17	& 68.44 & 68.58\\
			\textrm{S(1)} &	0 &	3.01 &	3.03 &	0.91 &	-0.01 &	-0.68\\
			\textrm{S(2)} &	1 &	2.76 &	2.78 &	0.41 &	-24.25 &	-25.01\\
			\textrm{S(3)} &	2 &	2.42 &	2.45 &	1.30 &	-40.12 &	-40.74\\
			\textrm{S(4)} &	3 &	2.17 &	2.20 & 	1.04 &	-46.19 &	-46.79\\
			\textrm{S(5)} &	4 &	2.02 &	2.04 &	0.85 &	-49.76 &	-50.32\\
			\textrm{S(6)} &	5 &	1.91 &	1.92 &	0.76 &	-52.17 &	-52.70\\
			\textrm{S(7)} &	6 &	1.82 &	1.83 &	0.64 &	-54.04 &    -54.56\\
			\textrm{S(8)} &	7 &	1.75 &	1.76 &	0.56 &	-55.42 &	-55.91\\
			\textrm{S(9)} &	8 &	1.69 &	1.70 &	0.53 &	-55.76 &	-56.24\\
			\textrm{S(10)} &	9 &	1.64 &	1.65 &	0.45 &	-56.73 &	-57.22\\
			\textrm{S(11)} &	10 & 1.61 &	1.61 &	0.43 &	-57.47 &	-57.98\\			
		\end{tabular}
	\end{ruledtabular}
\end{table}

\subsection{Dynamics of skyrmion bags in an anisotropy gradient with boundary potential}

\begin{figure}
	\includegraphics[width=1\columnwidth]{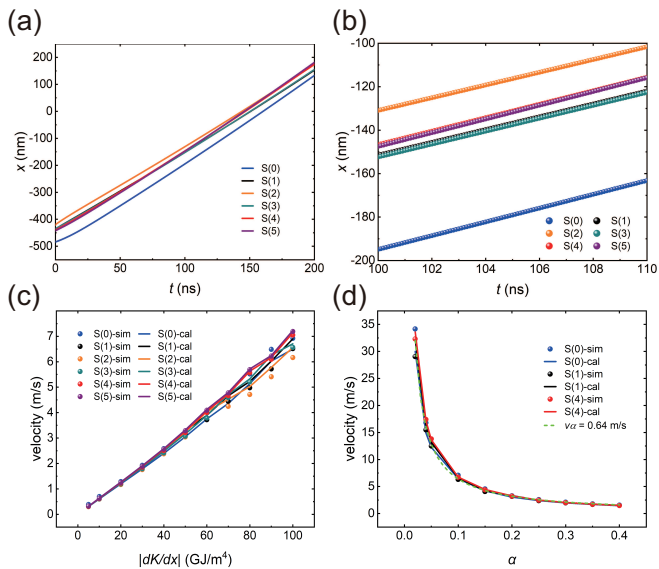}%
	\caption{\label{Fig:5} (a) The \textit{x}-coordinate of S(0)-S(5) as they move along the boundary versus time with $\alpha = 0.2$ and $\vert dK/dx \vert = 50\ \mathrm{GJ/m^4}$. (b) The enlarged view of (a) with the time from 100 to 110 ns. (c) The velocity of S(0)-S(5) as a function of  $\vert dK/dx \vert$. The dots are the simulated data, and the line is the data calculated from Eq.~(\ref{eq:9}). (d) The velocity of skyrmion bags as a function of $\alpha$ with $\vert dK/dx \vert = 50\ \mathrm{GJ/m^4}$. The green dotted line is an inverse function, $v\alpha = 0.64 \mathrm{m/s}$, where $v$ is velocity and $\alpha$ is damping constant.}
\end{figure}

For investigating the dynamics of skyrmion bags with boundary potential, the $L$ and $W$ of magnetic film are set to be 2048 nm and 256 nm, respectively. Due to the different maximum magnetic anisotropy constants, we put the skyrmion bags at $x = -450\ \mathrm{nm}$ except S(1) which is placed at $x = -300\ \mathrm{nm}$. In addition, due to the different signs of topological charge, S(0) is placed near the upper boundary, S(1) is placed in the middle of the film, and S(2) to S(5) are placed near the lower boundary. Figure~\ref{Fig:5}(a) shows the \textit{x}-coordinate of S(0)-S(5) as they move along the boundary versus time with $\alpha = 0.2$ and $\vert dK/dx \vert = 50\ \mathrm{GJ/m^4}$. Figure~\ref{Fig:5}(b) is an enlarged view of Fig.~\ref{Fig:5}(a) with the time from 100 to 110 ns, where the lines composed by dots are almost parallel, indicating that the skyrmion bags with different spin textures have the same velocity as they move along the boundary.

From Eq.~(\ref{eq:9}), it is found that the velocity of skyrmion bags is proportional to  $\vert dK/dx \vert$, while it is independent of $Q$. Figure~\ref{Fig:5}(c) shows the velocity of S(0)-S(5) as a function of $\vert dK/dx \vert$. The dots are the simulated data, and the line is the data calculated from Eq.~(\ref{eq:9}). It shows that the velocity of skyrmion bags linearly increases with the increasing $\vert dK/dx \vert$. It is worth noting that skyrmionium is generally considered to be faster than skyrmion\cite{R28,R46}. In our study, skyrmionium and skyrmion bags with arbitrary topological charge have the same velocity with boundary potential. When $\vert dK/dx \vert = 50\ \mathrm{GJ/m^4}$, the velocity of S(0)-S(5) is nearly 3.20 m/s. It is worth mentioning that by replacing the boundary with a material with a higher magnetic anisotropy constant, it will further improve the stability of skyrmion bags while moving along the boundary\cite{R47}. Next, we studied the velocity of S(0), S(1), and S(4) as a function of $\alpha$, as shown in Fig.~\ref{Fig:5}(d). It is found that with the increase of $\alpha$, the velocities of S(0), S(1), and S(4) gradually decrease and can be fitted by the green dotted line which is an inverse function, $v\alpha = 0.64\ \mathrm{m/s}$, where $v$ is velocity and $\alpha$ is damping constant. Furthermore, it is found that under the same $\alpha$, S(0), S(1), and S(4) have almost the same velocity, which indicates that the velocity of skyrmion bags has the same dependence on $\alpha$.

We also studied the velocity $v$, relative error parameter of velocity $\varepsilon_v$, term $u$, the component of dissipation tensor $\eta_{xx}$ and the value $u/\eta_{xx}$ of S(0) to S(11) with $\alpha = 0.2$ and $\vert dK/dx \vert = 50\ \mathrm{GJ/m^4}$ (see Table~\ref{tab:2}). It can be found that in the case of skyrmion bags with $\vert Q \vert > 2$, the $\varepsilon_v$ is less than 2.0 \%, which indicates that the simulation results are in good agreement with the analytically calculated results. For the skyrmion bags with $\vert Q \vert \leq 2$, the $\varepsilon_v$ is more than 2.0 \%. The reason is that small topological charge leads to the small Magnus force, which is not easy to balance with the force generated by the boundary potential, resulting in a velocity in $y$ direction. Hence, Eq.~(\ref{eq:9}) is not suitable to describe the dynamics of skyrmion bags with a small topological charge. Moreover, although both the $u$ and $\eta_{xx}$ are different for different skyrmion bags, the value of $u/\eta_{xx}$ is almost the same. When $\alpha = 0.2$ and $\vert dK/dx \vert = 50\ \mathrm{GJ/m^4}$, the $u/\eta_{xx}$ is about 4.20. From Eq.~(\ref{eq:9}), it can be understood that why the skyrmion bags with arbitrary topological charge in an anisotropy gradient have the same velocity when moving along the boundary.

\begin{table}[b]%The best place to locate the table environment is directly after its first reference in text
	\caption{\label{tab:2}%
		The topological charge $Q$, velocity $v$, relative error parameter of velocity $\varepsilon_v$, term $u$, component of dissipation tensor $\eta_{xx}$, and value $u/\eta_{xx}$ for each bag with  $\alpha = 0.2$ and $\vert dK/dx \vert = 50\ \mathrm{GJ/m^4}$.
	}
	\begin{ruledtabular}
		\begin{tabular}{cccccccc}
			\textrm{Bag}&
			\textit{Q}&
			\multicolumn{1}{c}{$\begin{aligned}&\ v_{sim}\\&\mathrm{(m/s)}\end{aligned}$}&
			\multicolumn{1}{c}{$\begin{aligned}&\ v_{cal}\\&\mathrm{(m/s)}\end{aligned}$}&
			\multicolumn{1}{c}{$\varepsilon_v\, \mathrm{(\%)}$}&
			\multicolumn{1}{c}{$\begin{aligned}&\quad u\\&\mathrm{(10^{-15})}\end{aligned}$}&
			\multicolumn{1}{c}{$\eta_{xx}$}&
			\multicolumn{1}{c}{$\begin{aligned}&\ u/\eta_{xx}\\&\mathrm{(10^{-17})}\end{aligned}$}\\
			\hline
			\textrm{S(0)} &	-1 &	3.28 &	3.05 &	6.96 &	0.93 &	23.08 &	4.02\\
			\textrm{S(1)} &	0 &	3.16 &	3.30 &	4.44 &	3.16 &	72.70 &	4.35\\
			\textrm{S(2)} &	1 &	3.04 &	3.24 &	6.53 &	4.54 &	106.13 &	4.27\\
			\textrm{S(3)} &	2 &	3.07 &	3.16 &	2.81 &	5.84 &	140.14	& 4.17\\
			\textrm{S(4)} &	3 &	3.22 &	3.27 &	1.55 &	7.15 &	166.03 &	4.31\\
			\textrm{S(5)} &	4 &	3.29 &	3.32 &	0.92 &	8.40 & 	191.96 &	4.38\\
			\textrm{S(6)} &	5 &	3.13 &	3.15 &	0.35 &	9.40 &	226.54	& 4.15\\
			\textrm{S(7)} &	6 &	3.15 &	3.16 &	0.28 &	10.45 &	250.66 &	4.17\\
			\textrm{S(8)} &	7 &	3.16 &	3.19 &	0.87 &	11.48 &	272.80 &	4.21\\
			\textrm{S(9)} &	8 &	3.23 &	3.25 &	0.53 &	12.58 &	293.96 &	4.28\\
			\textrm{S(10)} &	9 &	3.25 &	3.27 &	0.57 &	13.61 &	315.41 &	4.31\\
			\textrm{S(11)} &	10 &	3.26 &	3.29 &	1.00 &	14.56 &	335.79 &	4.34\\
						
		\end{tabular}
	\end{ruledtabular}
\end{table}

\section{Conclusion}
In conclusion, we first investigated the maximum magnetic anisotropy constant allowed for the stable existence of skyrmion bags. It is found that the maximum magnetic anisotropy constant of skyrmionium is smallest. Subsequently, we investigated the dynamics of skyrmion bags in an anisotropy gradient without boundary potential. In this case, the dynamics of skyrmion bags are found to be related to the spin textures. With the increase of $Q$, the velocity decreases and $\vert \theta_{Sky} \vert$ increases. With the increase of $\vert dK/dx \vert$, the velocity linearly increases and $\vert \theta_{Sky} \vert$ is unchanged. With the increase of $\alpha$, both the velocity and $\vert \theta_{Sky} \vert$ decrease. Moreover, the simulation results are in good agreement with the calculation results obtained by the Thiele equation. Although the skyrmion bags have arbitrary topological charge, there should be corresponding limits on the velocity and $\theta_{Sky}$. Finally, we investigated the dynamics of skyrmion bags in an anisotropy gradient with boundary potential, and found that with the increases of $Q$, the velocity is almost unchanged including for the skyrmionium, while the velocity linearly increases with the $\vert dK/dx \vert$ increasing. Additionally, with increasing $\alpha$, the velocity decreases and shows an inverse relationship. Moreover, it is found that in the presence of a boundary potential, the Thiele equation is not suitable to describe the dynamics of skyrmion bags with a small topological charge. Furthermore, it is found that although the term $u$ and $\eta_{xx}$ are related to the magnetization distribution of skyrmion bag, the value of $u/\eta_{xx}$ is almost the same and results in the same velocity of skyrmion bags when moving along the boundary. Our results about the skyrmion bag dynamics in an anisotropy gradient can play an important role in promoting the generation and application of racetrack memory based on skyrmion bags.

\begin{acknowledgments}
This work was supported by the National Natural Science Foundation of China (Grants No. 12074158, No. 12174166 and No. 12104197).
\end{acknowledgments}
\nocite{*}
\bibliography{reference}

\begin{thebibliography}{49}%
\makeatletter
\providecommand \@ifxundefined [1]{%
 \@ifx{#1\undefined}
}%
\providecommand \@ifnum [1]{%
 \ifnum #1\expandafter \@firstoftwo
 \else \expandafter \@secondoftwo
 \fi
}%
\providecommand \@ifx [1]{%
 \ifx #1\expandafter \@firstoftwo
 \else \expandafter \@secondoftwo
 \fi
}%
\providecommand \natexlab [1]{#1}%
\providecommand \enquote  [1]{``#1''}%
\providecommand \bibnamefont  [1]{#1}%
\providecommand \bibfnamefont [1]{#1}%
\providecommand \citenamefont [1]{#1}%
\providecommand \href@noop [0]{\@secondoftwo}%
\providecommand \href [0]{\begingroup \@sanitize@url \@href}%
\providecommand \@href[1]{\@@startlink{#1}\@@href}%
\providecommand \@@href[1]{\endgroup#1\@@endlink}%
\providecommand \@sanitize@url [0]{\catcode `\\12\catcode `\$12\catcode
  `\&12\catcode `\#12\catcode `\^12\catcode `\_12\catcode `\%12\relax}%
\providecommand \@@startlink[1]{}%
\providecommand \@@endlink[0]{}%
\providecommand \url  [0]{\begingroup\@sanitize@url \@url }%
\providecommand \@url [1]{\endgroup\@href {#1}{\urlprefix }}%
\providecommand \urlprefix  [0]{URL }%
\providecommand \Eprint [0]{\href }%
\providecommand \doibase [0]{https://doi.org/}%
\providecommand \selectlanguage [0]{\@gobble}%
\providecommand \bibinfo  [0]{\@secondoftwo}%
\providecommand \bibfield  [0]{\@secondoftwo}%
\providecommand \translation [1]{[#1]}%
\providecommand \BibitemOpen [0]{}%
\providecommand \bibitemStop [0]{}%
\providecommand \bibitemNoStop [0]{.\EOS\space}%
\providecommand \EOS [0]{\spacefactor3000\relax}%
\providecommand \BibitemShut  [1]{\csname bibitem#1\endcsname}%
\let\auto@bib@innerbib\@empty
%</preamble>
\bibitem [{\citenamefont {Mühlbauer}\ \emph {et~al.}(2009)\citenamefont
  {Mühlbauer}, \citenamefont {Binz}, \citenamefont {Jonietz}, \citenamefont
  {Pfleiderer}, \citenamefont {Rosch}, \citenamefont {Neubauer}, \citenamefont
  {Georgii},\ and\ \citenamefont {Böni}}]{R1}%
  \BibitemOpen
  \bibfield  {author} {\bibinfo {author} {\bibfnamefont {S.}~\bibnamefont
  {Mühlbauer}}, \bibinfo {author} {\bibfnamefont {B.}~\bibnamefont {Binz}},
  \bibinfo {author} {\bibfnamefont {F.}~\bibnamefont {Jonietz}}, \bibinfo
  {author} {\bibfnamefont {C.}~\bibnamefont {Pfleiderer}}, \bibinfo {author}
  {\bibfnamefont {A.}~\bibnamefont {Rosch}}, \bibinfo {author} {\bibfnamefont
  {A.}~\bibnamefont {Neubauer}}, \bibinfo {author} {\bibfnamefont
  {R.}~\bibnamefont {Georgii}},\ and\ \bibinfo {author} {\bibfnamefont
  {P.}~\bibnamefont {Böni}},\ }\href@noop {} {\bibfield  {journal} {\bibinfo
  {journal} {Science}\ }\textbf {\bibinfo {volume} {323}},\ \bibinfo {pages}
  {915} (\bibinfo {year} {2009})}\BibitemShut {NoStop}%
\bibitem [{\citenamefont {Yu}\ \emph {et~al.}(2010)\citenamefont {Yu},
  \citenamefont {Onose}, \citenamefont {Kanazawa}, \citenamefont {Park},
  \citenamefont {Han}, \citenamefont {Matsui}, \citenamefont {Nagaosa},\ and\
  \citenamefont {Tokura}}]{R2}%
  \BibitemOpen
  \bibfield  {author} {\bibinfo {author} {\bibfnamefont {X.}~\bibnamefont
  {Yu}}, \bibinfo {author} {\bibfnamefont {Y.}~\bibnamefont {Onose}}, \bibinfo
  {author} {\bibfnamefont {N.}~\bibnamefont {Kanazawa}}, \bibinfo {author}
  {\bibfnamefont {J.~H.}\ \bibnamefont {Park}}, \bibinfo {author}
  {\bibfnamefont {J.}~\bibnamefont {Han}}, \bibinfo {author} {\bibfnamefont
  {Y.}~\bibnamefont {Matsui}}, \bibinfo {author} {\bibfnamefont
  {N.}~\bibnamefont {Nagaosa}},\ and\ \bibinfo {author} {\bibfnamefont
  {Y.}~\bibnamefont {Tokura}},\ }\href@noop {} {\bibfield  {journal} {\bibinfo
  {journal} {Nature}\ }\textbf {\bibinfo {volume} {465}},\ \bibinfo {pages}
  {901} (\bibinfo {year} {2010})}\BibitemShut {NoStop}%
\bibitem [{\citenamefont {Sampaio}\ \emph {et~al.}(2013)\citenamefont
  {Sampaio}, \citenamefont {Cros}, \citenamefont {Rohart}, \citenamefont
  {Thiaville},\ and\ \citenamefont {Fert}}]{R3}%
  \BibitemOpen
  \bibfield  {author} {\bibinfo {author} {\bibfnamefont {J.}~\bibnamefont
  {Sampaio}}, \bibinfo {author} {\bibfnamefont {V.}~\bibnamefont {Cros}},
  \bibinfo {author} {\bibfnamefont {S.}~\bibnamefont {Rohart}}, \bibinfo
  {author} {\bibfnamefont {A.}~\bibnamefont {Thiaville}},\ and\ \bibinfo
  {author} {\bibfnamefont {A.}~\bibnamefont {Fert}},\ }\href@noop {} {\bibfield
   {journal} {\bibinfo  {journal} {Nature nanotechnology}\ }\textbf {\bibinfo
  {volume} {8}},\ \bibinfo {pages} {839} (\bibinfo {year} {2013})}\BibitemShut
  {NoStop}%
\bibitem [{\citenamefont {Fert}\ \emph {et~al.}(2013)\citenamefont {Fert},
  \citenamefont {Cros},\ and\ \citenamefont {Sampaio}}]{R4}%
  \BibitemOpen
  \bibfield  {author} {\bibinfo {author} {\bibfnamefont {A.}~\bibnamefont
  {Fert}}, \bibinfo {author} {\bibfnamefont {V.}~\bibnamefont {Cros}},\ and\
  \bibinfo {author} {\bibfnamefont {J.}~\bibnamefont {Sampaio}},\ }\href@noop
  {} {\bibfield  {journal} {\bibinfo  {journal} {Nature nanotechnology}\
  }\textbf {\bibinfo {volume} {8}},\ \bibinfo {pages} {152} (\bibinfo {year}
  {2013})}\BibitemShut {NoStop}%
\bibitem [{\citenamefont {Tomasello}\ \emph {et~al.}(2014)\citenamefont
  {Tomasello}, \citenamefont {Martinez}, \citenamefont {Zivieri}, \citenamefont
  {Torres}, \citenamefont {Carpentieri},\ and\ \citenamefont {Finocchio}}]{R5}%
  \BibitemOpen
  \bibfield  {author} {\bibinfo {author} {\bibfnamefont {R.}~\bibnamefont
  {Tomasello}}, \bibinfo {author} {\bibfnamefont {E.}~\bibnamefont {Martinez}},
  \bibinfo {author} {\bibfnamefont {R.}~\bibnamefont {Zivieri}}, \bibinfo
  {author} {\bibfnamefont {L.}~\bibnamefont {Torres}}, \bibinfo {author}
  {\bibfnamefont {M.}~\bibnamefont {Carpentieri}},\ and\ \bibinfo {author}
  {\bibfnamefont {G.}~\bibnamefont {Finocchio}},\ }\href@noop {} {\bibfield
  {journal} {\bibinfo  {journal} {Scientific reports}\ }\textbf {\bibinfo
  {volume} {4}},\ \bibinfo {pages} {1} (\bibinfo {year} {2014})}\BibitemShut
  {NoStop}%
\bibitem [{\citenamefont {Zhang}\ \emph
  {et~al.}(2015{\natexlab{a}})\citenamefont {Zhang}, \citenamefont {Ezawa},\
  and\ \citenamefont {Zhou}}]{R6}%
  \BibitemOpen
  \bibfield  {author} {\bibinfo {author} {\bibfnamefont {X.}~\bibnamefont
  {Zhang}}, \bibinfo {author} {\bibfnamefont {M.}~\bibnamefont {Ezawa}},\ and\
  \bibinfo {author} {\bibfnamefont {Y.}~\bibnamefont {Zhou}},\ }\href@noop {}
  {\bibfield  {journal} {\bibinfo  {journal} {Scientific reports}\ }\textbf
  {\bibinfo {volume} {5}},\ \bibinfo {pages} {1} (\bibinfo {year}
  {2015}{\natexlab{a}})}\BibitemShut {NoStop}%
\bibitem [{\citenamefont {Luo}\ \emph {et~al.}(2018)\citenamefont {Luo},
  \citenamefont {Song}, \citenamefont {Li}, \citenamefont {Zhang},
  \citenamefont {Hong}, \citenamefont {Yang}, \citenamefont {Zou},
  \citenamefont {Xu},\ and\ \citenamefont {You}}]{R7}%
  \BibitemOpen
  \bibfield  {author} {\bibinfo {author} {\bibfnamefont {S.}~\bibnamefont
  {Luo}}, \bibinfo {author} {\bibfnamefont {M.}~\bibnamefont {Song}}, \bibinfo
  {author} {\bibfnamefont {X.}~\bibnamefont {Li}}, \bibinfo {author}
  {\bibfnamefont {Y.}~\bibnamefont {Zhang}}, \bibinfo {author} {\bibfnamefont
  {J.}~\bibnamefont {Hong}}, \bibinfo {author} {\bibfnamefont {X.}~\bibnamefont
  {Yang}}, \bibinfo {author} {\bibfnamefont {X.}~\bibnamefont {Zou}}, \bibinfo
  {author} {\bibfnamefont {N.}~\bibnamefont {Xu}},\ and\ \bibinfo {author}
  {\bibfnamefont {L.}~\bibnamefont {You}},\ }\href@noop {} {\bibfield
  {journal} {\bibinfo  {journal} {Nano letters}\ }\textbf {\bibinfo {volume}
  {18}},\ \bibinfo {pages} {1180} (\bibinfo {year} {2018})}\BibitemShut
  {NoStop}%
\bibitem [{\citenamefont {Luo}\ and\ \citenamefont {You}(2021)}]{R8}%
  \BibitemOpen
  \bibfield  {author} {\bibinfo {author} {\bibfnamefont {S.}~\bibnamefont
  {Luo}}\ and\ \bibinfo {author} {\bibfnamefont {L.}~\bibnamefont {You}},\
  }\href@noop {} {\bibfield  {journal} {\bibinfo  {journal} {APL Materials}\
  }\textbf {\bibinfo {volume} {9}},\ \bibinfo {pages} {050901} (\bibinfo {year}
  {2021})}\BibitemShut {NoStop}%
\bibitem [{\citenamefont {Huang}\ \emph {et~al.}(2017)\citenamefont {Huang},
  \citenamefont {Kang}, \citenamefont {Zhang}, \citenamefont {Zhou},\ and\
  \citenamefont {Zhao}}]{R9}%
  \BibitemOpen
  \bibfield  {author} {\bibinfo {author} {\bibfnamefont {Y.}~\bibnamefont
  {Huang}}, \bibinfo {author} {\bibfnamefont {W.}~\bibnamefont {Kang}},
  \bibinfo {author} {\bibfnamefont {X.}~\bibnamefont {Zhang}}, \bibinfo
  {author} {\bibfnamefont {Y.}~\bibnamefont {Zhou}},\ and\ \bibinfo {author}
  {\bibfnamefont {W.}~\bibnamefont {Zhao}},\ }\href@noop {} {\bibfield
  {journal} {\bibinfo  {journal} {Nanotechnology}\ }\textbf {\bibinfo {volume}
  {28}},\ \bibinfo {pages} {08LT02} (\bibinfo {year} {2017})}\BibitemShut
  {NoStop}%
\bibitem [{\citenamefont {Li}\ \emph {et~al.}(2017)\citenamefont {Li},
  \citenamefont {Kang}, \citenamefont {Huang}, \citenamefont {Zhang},
  \citenamefont {Zhou},\ and\ \citenamefont {Zhao}}]{R10}%
  \BibitemOpen
  \bibfield  {author} {\bibinfo {author} {\bibfnamefont {S.}~\bibnamefont
  {Li}}, \bibinfo {author} {\bibfnamefont {W.}~\bibnamefont {Kang}}, \bibinfo
  {author} {\bibfnamefont {Y.}~\bibnamefont {Huang}}, \bibinfo {author}
  {\bibfnamefont {X.}~\bibnamefont {Zhang}}, \bibinfo {author} {\bibfnamefont
  {Y.}~\bibnamefont {Zhou}},\ and\ \bibinfo {author} {\bibfnamefont
  {W.}~\bibnamefont {Zhao}},\ }\href@noop {} {\bibfield  {journal} {\bibinfo
  {journal} {Nanotechnology}\ }\textbf {\bibinfo {volume} {28}},\ \bibinfo
  {pages} {31LT01} (\bibinfo {year} {2017})}\BibitemShut {NoStop}%
\bibitem [{\citenamefont {Chen}\ \emph {et~al.}(2018)\citenamefont {Chen},
  \citenamefont {Kang}, \citenamefont {Zhu}, \citenamefont {Zhang},
  \citenamefont {Lei}, \citenamefont {Zhang}, \citenamefont {Zhou},\ and\
  \citenamefont {Zhao}}]{R11}%
  \BibitemOpen
  \bibfield  {author} {\bibinfo {author} {\bibfnamefont {X.}~\bibnamefont
  {Chen}}, \bibinfo {author} {\bibfnamefont {W.}~\bibnamefont {Kang}}, \bibinfo
  {author} {\bibfnamefont {D.}~\bibnamefont {Zhu}}, \bibinfo {author}
  {\bibfnamefont {X.}~\bibnamefont {Zhang}}, \bibinfo {author} {\bibfnamefont
  {N.}~\bibnamefont {Lei}}, \bibinfo {author} {\bibfnamefont {Y.}~\bibnamefont
  {Zhang}}, \bibinfo {author} {\bibfnamefont {Y.}~\bibnamefont {Zhou}},\ and\
  \bibinfo {author} {\bibfnamefont {W.}~\bibnamefont {Zhao}},\ }\href@noop {}
  {\bibfield  {journal} {\bibinfo  {journal} {Nanoscale}\ }\textbf {\bibinfo
  {volume} {10}},\ \bibinfo {pages} {6139} (\bibinfo {year}
  {2018})}\BibitemShut {NoStop}%
\bibitem [{\citenamefont {Iwasaki}\ \emph
  {et~al.}(2013{\natexlab{a}})\citenamefont {Iwasaki}, \citenamefont
  {Mochizuki},\ and\ \citenamefont {Nagaosa}}]{R12}%
  \BibitemOpen
  \bibfield  {author} {\bibinfo {author} {\bibfnamefont {J.}~\bibnamefont
  {Iwasaki}}, \bibinfo {author} {\bibfnamefont {M.}~\bibnamefont {Mochizuki}},\
  and\ \bibinfo {author} {\bibfnamefont {N.}~\bibnamefont {Nagaosa}},\
  }\href@noop {} {\bibfield  {journal} {\bibinfo  {journal} {Nature
  communications}\ }\textbf {\bibinfo {volume} {4}},\ \bibinfo {pages} {1}
  (\bibinfo {year} {2013}{\natexlab{a}})}\BibitemShut {NoStop}%
\bibitem [{\citenamefont {Iwasaki}\ \emph
  {et~al.}(2013{\natexlab{b}})\citenamefont {Iwasaki}, \citenamefont
  {Mochizuki},\ and\ \citenamefont {Nagaosa}}]{R13}%
  \BibitemOpen
  \bibfield  {author} {\bibinfo {author} {\bibfnamefont {J.}~\bibnamefont
  {Iwasaki}}, \bibinfo {author} {\bibfnamefont {M.}~\bibnamefont {Mochizuki}},\
  and\ \bibinfo {author} {\bibfnamefont {N.}~\bibnamefont {Nagaosa}},\
  }\href@noop {} {\bibfield  {journal} {\bibinfo  {journal} {Nature
  nanotechnology}\ }\textbf {\bibinfo {volume} {8}},\ \bibinfo {pages} {742}
  (\bibinfo {year} {2013}{\natexlab{b}})}\BibitemShut {NoStop}%
\bibitem [{\citenamefont {Yu}\ \emph {et~al.}(2012)\citenamefont {Yu},
  \citenamefont {Kanazawa}, \citenamefont {Zhang}, \citenamefont {Nagai},
  \citenamefont {Hara}, \citenamefont {Kimoto}, \citenamefont {Matsui},
  \citenamefont {Onose},\ and\ \citenamefont {Tokura}}]{R14}%
  \BibitemOpen
  \bibfield  {author} {\bibinfo {author} {\bibfnamefont {X.}~\bibnamefont
  {Yu}}, \bibinfo {author} {\bibfnamefont {N.}~\bibnamefont {Kanazawa}},
  \bibinfo {author} {\bibfnamefont {W.}~\bibnamefont {Zhang}}, \bibinfo
  {author} {\bibfnamefont {T.}~\bibnamefont {Nagai}}, \bibinfo {author}
  {\bibfnamefont {T.}~\bibnamefont {Hara}}, \bibinfo {author} {\bibfnamefont
  {K.}~\bibnamefont {Kimoto}}, \bibinfo {author} {\bibfnamefont
  {Y.}~\bibnamefont {Matsui}}, \bibinfo {author} {\bibfnamefont
  {Y.}~\bibnamefont {Onose}},\ and\ \bibinfo {author} {\bibfnamefont
  {Y.}~\bibnamefont {Tokura}},\ }\href@noop {} {\bibfield  {journal} {\bibinfo
  {journal} {Nature communications}\ }\textbf {\bibinfo {volume} {3}},\
  \bibinfo {pages} {1} (\bibinfo {year} {2012})}\BibitemShut {NoStop}%
\bibitem [{\citenamefont {G{\"o}bel}\ \emph {et~al.}(2019)\citenamefont
  {G{\"o}bel}, \citenamefont {Mook}, \citenamefont {Henk},\ and\ \citenamefont
  {Mertig}}]{R15}%
  \BibitemOpen
  \bibfield  {author} {\bibinfo {author} {\bibfnamefont {B.}~\bibnamefont
  {G{\"o}bel}}, \bibinfo {author} {\bibfnamefont {A.}~\bibnamefont {Mook}},
  \bibinfo {author} {\bibfnamefont {J.}~\bibnamefont {Henk}},\ and\ \bibinfo
  {author} {\bibfnamefont {I.}~\bibnamefont {Mertig}},\ }\href@noop {}
  {\bibfield  {journal} {\bibinfo  {journal} {Physical Review B}\ }\textbf
  {\bibinfo {volume} {99}},\ \bibinfo {pages} {020405} (\bibinfo {year}
  {2019})}\BibitemShut {NoStop}%
\bibitem [{\citenamefont {Zhang}\ \emph
  {et~al.}(2015{\natexlab{b}})\citenamefont {Zhang}, \citenamefont {Ezawa},
  \citenamefont {Xiao}, \citenamefont {Zhao}, \citenamefont {Liu},\ and\
  \citenamefont {Zhou}}]{R16}%
  \BibitemOpen
  \bibfield  {author} {\bibinfo {author} {\bibfnamefont {X.}~\bibnamefont
  {Zhang}}, \bibinfo {author} {\bibfnamefont {M.}~\bibnamefont {Ezawa}},
  \bibinfo {author} {\bibfnamefont {D.}~\bibnamefont {Xiao}}, \bibinfo {author}
  {\bibfnamefont {G.}~\bibnamefont {Zhao}}, \bibinfo {author} {\bibfnamefont
  {Y.}~\bibnamefont {Liu}},\ and\ \bibinfo {author} {\bibfnamefont
  {Y.}~\bibnamefont {Zhou}},\ }\href@noop {} {\bibfield  {journal} {\bibinfo
  {journal} {Nanotechnology}\ }\textbf {\bibinfo {volume} {26}},\ \bibinfo
  {pages} {225701} (\bibinfo {year} {2015}{\natexlab{b}})}\BibitemShut
  {NoStop}%
\bibitem [{\citenamefont {Iwasaki}\ \emph {et~al.}(2014)\citenamefont
  {Iwasaki}, \citenamefont {Beekman},\ and\ \citenamefont {Nagaosa}}]{R17}%
  \BibitemOpen
  \bibfield  {author} {\bibinfo {author} {\bibfnamefont {J.}~\bibnamefont
  {Iwasaki}}, \bibinfo {author} {\bibfnamefont {A.~J.}\ \bibnamefont
  {Beekman}},\ and\ \bibinfo {author} {\bibfnamefont {N.}~\bibnamefont
  {Nagaosa}},\ }\href@noop {} {\bibfield  {journal} {\bibinfo  {journal}
  {Physical Review B}\ }\textbf {\bibinfo {volume} {89}},\ \bibinfo {pages}
  {064412} (\bibinfo {year} {2014})}\BibitemShut {NoStop}%
\bibitem [{\citenamefont {Wang}\ \emph {et~al.}(2017)\citenamefont {Wang},
  \citenamefont {Xiao}, \citenamefont {Chen}, \citenamefont {Zhou},\ and\
  \citenamefont {Liu}}]{R18}%
  \BibitemOpen
  \bibfield  {author} {\bibinfo {author} {\bibfnamefont {C.}~\bibnamefont
  {Wang}}, \bibinfo {author} {\bibfnamefont {D.}~\bibnamefont {Xiao}}, \bibinfo
  {author} {\bibfnamefont {X.}~\bibnamefont {Chen}}, \bibinfo {author}
  {\bibfnamefont {Y.}~\bibnamefont {Zhou}},\ and\ \bibinfo {author}
  {\bibfnamefont {Y.}~\bibnamefont {Liu}},\ }\href@noop {} {\bibfield
  {journal} {\bibinfo  {journal} {New Journal of Physics}\ }\textbf {\bibinfo
  {volume} {19}},\ \bibinfo {pages} {083008} (\bibinfo {year}
  {2017})}\BibitemShut {NoStop}%
\bibitem [{\citenamefont {Zhang}\ \emph {et~al.}(2018)\citenamefont {Zhang},
  \citenamefont {Wang}, \citenamefont {Burn}, \citenamefont {Peng},
  \citenamefont {Berger}, \citenamefont {Bauer}, \citenamefont {Pfleiderer},
  \citenamefont {Van Der~Laan},\ and\ \citenamefont {Hesjedal}}]{R19}%
  \BibitemOpen
  \bibfield  {author} {\bibinfo {author} {\bibfnamefont {S.}~\bibnamefont
  {Zhang}}, \bibinfo {author} {\bibfnamefont {W.}~\bibnamefont {Wang}},
  \bibinfo {author} {\bibfnamefont {D.}~\bibnamefont {Burn}}, \bibinfo {author}
  {\bibfnamefont {H.}~\bibnamefont {Peng}}, \bibinfo {author} {\bibfnamefont
  {H.}~\bibnamefont {Berger}}, \bibinfo {author} {\bibfnamefont
  {A.}~\bibnamefont {Bauer}}, \bibinfo {author} {\bibfnamefont
  {C.}~\bibnamefont {Pfleiderer}}, \bibinfo {author} {\bibfnamefont
  {G.}~\bibnamefont {Van Der~Laan}},\ and\ \bibinfo {author} {\bibfnamefont
  {T.}~\bibnamefont {Hesjedal}},\ }\href@noop {} {\bibfield  {journal}
  {\bibinfo  {journal} {Nature communications}\ }\textbf {\bibinfo {volume}
  {9}},\ \bibinfo {pages} {1} (\bibinfo {year} {2018})}\BibitemShut {NoStop}%
\bibitem [{\citenamefont {Kong}\ and\ \citenamefont {Zang}(2013)}]{R20}%
  \BibitemOpen
  \bibfield  {author} {\bibinfo {author} {\bibfnamefont {L.}~\bibnamefont
  {Kong}}\ and\ \bibinfo {author} {\bibfnamefont {J.}~\bibnamefont {Zang}},\
  }\href@noop {} {\bibfield  {journal} {\bibinfo  {journal} {Physical review
  letters}\ }\textbf {\bibinfo {volume} {111}},\ \bibinfo {pages} {067203}
  (\bibinfo {year} {2013})}\BibitemShut {NoStop}%
\bibitem [{\citenamefont {Mochizuki}\ \emph {et~al.}(2014)\citenamefont
  {Mochizuki}, \citenamefont {Yu}, \citenamefont {Seki}, \citenamefont
  {Kanazawa}, \citenamefont {Koshibae}, \citenamefont {Zang}, \citenamefont
  {Mostovoy}, \citenamefont {Tokura},\ and\ \citenamefont {Nagaosa}}]{R21}%
  \BibitemOpen
  \bibfield  {author} {\bibinfo {author} {\bibfnamefont {M.}~\bibnamefont
  {Mochizuki}}, \bibinfo {author} {\bibfnamefont {X.}~\bibnamefont {Yu}},
  \bibinfo {author} {\bibfnamefont {S.}~\bibnamefont {Seki}}, \bibinfo {author}
  {\bibfnamefont {N.}~\bibnamefont {Kanazawa}}, \bibinfo {author}
  {\bibfnamefont {W.}~\bibnamefont {Koshibae}}, \bibinfo {author}
  {\bibfnamefont {J.}~\bibnamefont {Zang}}, \bibinfo {author} {\bibfnamefont
  {M.}~\bibnamefont {Mostovoy}}, \bibinfo {author} {\bibfnamefont
  {Y.}~\bibnamefont {Tokura}},\ and\ \bibinfo {author} {\bibfnamefont
  {N.}~\bibnamefont {Nagaosa}},\ }\href@noop {} {\bibfield  {journal} {\bibinfo
   {journal} {Nature materials}\ }\textbf {\bibinfo {volume} {13}},\ \bibinfo
  {pages} {241} (\bibinfo {year} {2014})}\BibitemShut {NoStop}%
\bibitem [{\citenamefont {Wang}\ \emph {et~al.}(2015)\citenamefont {Wang},
  \citenamefont {Beg}, \citenamefont {Zhang}, \citenamefont {Kuch},\ and\
  \citenamefont {Fangohr}}]{R22}%
  \BibitemOpen
  \bibfield  {author} {\bibinfo {author} {\bibfnamefont {W.}~\bibnamefont
  {Wang}}, \bibinfo {author} {\bibfnamefont {M.}~\bibnamefont {Beg}}, \bibinfo
  {author} {\bibfnamefont {B.}~\bibnamefont {Zhang}}, \bibinfo {author}
  {\bibfnamefont {W.}~\bibnamefont {Kuch}},\ and\ \bibinfo {author}
  {\bibfnamefont {H.}~\bibnamefont {Fangohr}},\ }\href@noop {} {\bibfield
  {journal} {\bibinfo  {journal} {Physical Review B}\ }\textbf {\bibinfo
  {volume} {92}},\ \bibinfo {pages} {020403} (\bibinfo {year}
  {2015})}\BibitemShut {NoStop}%
\bibitem [{\citenamefont {Ikka}\ \emph {et~al.}(2018)\citenamefont {Ikka},
  \citenamefont {Takeuchi},\ and\ \citenamefont {Mochizuki}}]{R23}%
  \BibitemOpen
  \bibfield  {author} {\bibinfo {author} {\bibfnamefont {M.}~\bibnamefont
  {Ikka}}, \bibinfo {author} {\bibfnamefont {A.}~\bibnamefont {Takeuchi}},\
  and\ \bibinfo {author} {\bibfnamefont {M.}~\bibnamefont {Mochizuki}},\
  }\href@noop {} {\bibfield  {journal} {\bibinfo  {journal} {Physical Review
  B}\ }\textbf {\bibinfo {volume} {98}},\ \bibinfo {pages} {184428} (\bibinfo
  {year} {2018})}\BibitemShut {NoStop}%
\bibitem [{\citenamefont {Weisheit}\ \emph {et~al.}(2007)\citenamefont
  {Weisheit}, \citenamefont {F{\~A}hler}, \citenamefont {Marty}, \citenamefont
  {Souche}, \citenamefont {Poinsignon},\ and\ \citenamefont {Givord}}]{R24}%
  \BibitemOpen
  \bibfield  {author} {\bibinfo {author} {\bibfnamefont {M.}~\bibnamefont
  {Weisheit}}, \bibinfo {author} {\bibfnamefont {S.}~\bibnamefont
  {F{\~A}hler}}, \bibinfo {author} {\bibfnamefont {A.}~\bibnamefont {Marty}},
  \bibinfo {author} {\bibfnamefont {Y.}~\bibnamefont {Souche}}, \bibinfo
  {author} {\bibfnamefont {C.}~\bibnamefont {Poinsignon}},\ and\ \bibinfo
  {author} {\bibfnamefont {D.}~\bibnamefont {Givord}},\ }\href@noop {}
  {\bibfield  {journal} {\bibinfo  {journal} {Science}\ }\textbf {\bibinfo
  {volume} {315}},\ \bibinfo {pages} {349} (\bibinfo {year}
  {2007})}\BibitemShut {NoStop}%
\bibitem [{\citenamefont {Maruyama}\ \emph {et~al.}(2009)\citenamefont
  {Maruyama}, \citenamefont {Shiota}, \citenamefont {Nozaki}, \citenamefont
  {Ohta}, \citenamefont {Toda}, \citenamefont {Mizuguchi}, \citenamefont
  {Tulapurkar}, \citenamefont {Shinjo}, \citenamefont {Shiraishi},
  \citenamefont {Mizukami} \emph {et~al.}}]{R25}%
  \BibitemOpen
  \bibfield  {author} {\bibinfo {author} {\bibfnamefont {T.}~\bibnamefont
  {Maruyama}}, \bibinfo {author} {\bibfnamefont {Y.}~\bibnamefont {Shiota}},
  \bibinfo {author} {\bibfnamefont {T.}~\bibnamefont {Nozaki}}, \bibinfo
  {author} {\bibfnamefont {K.}~\bibnamefont {Ohta}}, \bibinfo {author}
  {\bibfnamefont {N.}~\bibnamefont {Toda}}, \bibinfo {author} {\bibfnamefont
  {M.}~\bibnamefont {Mizuguchi}}, \bibinfo {author} {\bibfnamefont
  {A.}~\bibnamefont {Tulapurkar}}, \bibinfo {author} {\bibfnamefont
  {T.}~\bibnamefont {Shinjo}}, \bibinfo {author} {\bibfnamefont
  {M.}~\bibnamefont {Shiraishi}}, \bibinfo {author} {\bibfnamefont
  {S.}~\bibnamefont {Mizukami}}, \emph {et~al.},\ }\href@noop {} {\bibfield
  {journal} {\bibinfo  {journal} {Nature nanotechnology}\ }\textbf {\bibinfo
  {volume} {4}},\ \bibinfo {pages} {158} (\bibinfo {year} {2009})}\BibitemShut
  {NoStop}%
\bibitem [{\citenamefont {Xia}\ \emph {et~al.}(2018)\citenamefont {Xia},
  \citenamefont {Song}, \citenamefont {Jin}, \citenamefont {Wang},
  \citenamefont {Wang},\ and\ \citenamefont {Liu}}]{R26}%
  \BibitemOpen
  \bibfield  {author} {\bibinfo {author} {\bibfnamefont {H.}~\bibnamefont
  {Xia}}, \bibinfo {author} {\bibfnamefont {C.}~\bibnamefont {Song}}, \bibinfo
  {author} {\bibfnamefont {C.}~\bibnamefont {Jin}}, \bibinfo {author}
  {\bibfnamefont {J.}~\bibnamefont {Wang}}, \bibinfo {author} {\bibfnamefont
  {J.}~\bibnamefont {Wang}},\ and\ \bibinfo {author} {\bibfnamefont
  {Q.}~\bibnamefont {Liu}},\ }\href@noop {} {\bibfield  {journal} {\bibinfo
  {journal} {Journal of Magnetism and Magnetic Materials}\ }\textbf {\bibinfo
  {volume} {458}},\ \bibinfo {pages} {57} (\bibinfo {year} {2018})}\BibitemShut
  {NoStop}%
\bibitem [{\citenamefont {Shen}\ \emph {et~al.}(2018)\citenamefont {Shen},
  \citenamefont {Xia}, \citenamefont {Zhao}, \citenamefont {Zhang},
  \citenamefont {Ezawa}, \citenamefont {Tretiakov}, \citenamefont {Liu},\ and\
  \citenamefont {Zhou}}]{R27}%
  \BibitemOpen
  \bibfield  {author} {\bibinfo {author} {\bibfnamefont {L.}~\bibnamefont
  {Shen}}, \bibinfo {author} {\bibfnamefont {J.}~\bibnamefont {Xia}}, \bibinfo
  {author} {\bibfnamefont {G.}~\bibnamefont {Zhao}}, \bibinfo {author}
  {\bibfnamefont {X.}~\bibnamefont {Zhang}}, \bibinfo {author} {\bibfnamefont
  {M.}~\bibnamefont {Ezawa}}, \bibinfo {author} {\bibfnamefont {O.~A.}\
  \bibnamefont {Tretiakov}}, \bibinfo {author} {\bibfnamefont {X.}~\bibnamefont
  {Liu}},\ and\ \bibinfo {author} {\bibfnamefont {Y.}~\bibnamefont {Zhou}},\
  }\href@noop {} {\bibfield  {journal} {\bibinfo  {journal} {Physical Review
  B}\ }\textbf {\bibinfo {volume} {98}},\ \bibinfo {pages} {134448} (\bibinfo
  {year} {2018})}\BibitemShut {NoStop}%
\bibitem [{\citenamefont {Song}\ \emph {et~al.}(2019)\citenamefont {Song},
  \citenamefont {Jin}, \citenamefont {Wang}, \citenamefont {Ma}, \citenamefont
  {Xia}, \citenamefont {Wang}, \citenamefont {Wang},\ and\ \citenamefont
  {Liu}}]{R28}%
  \BibitemOpen
  \bibfield  {author} {\bibinfo {author} {\bibfnamefont {C.}~\bibnamefont
  {Song}}, \bibinfo {author} {\bibfnamefont {C.}~\bibnamefont {Jin}}, \bibinfo
  {author} {\bibfnamefont {J.}~\bibnamefont {Wang}}, \bibinfo {author}
  {\bibfnamefont {Y.}~\bibnamefont {Ma}}, \bibinfo {author} {\bibfnamefont
  {H.}~\bibnamefont {Xia}}, \bibinfo {author} {\bibfnamefont {J.}~\bibnamefont
  {Wang}}, \bibinfo {author} {\bibfnamefont {J.}~\bibnamefont {Wang}},\ and\
  \bibinfo {author} {\bibfnamefont {Q.}~\bibnamefont {Liu}},\ }\href@noop {}
  {\bibfield  {journal} {\bibinfo  {journal} {Applied Physics Express}\
  }\textbf {\bibinfo {volume} {12}},\ \bibinfo {pages} {083003} (\bibinfo
  {year} {2019})}\BibitemShut {NoStop}%
\bibitem [{\citenamefont {Tomasello}\ \emph {et~al.}(2018)\citenamefont
  {Tomasello}, \citenamefont {Komineas}, \citenamefont {Siracusano},
  \citenamefont {Carpentieri},\ and\ \citenamefont {Finocchio}}]{R29}%
  \BibitemOpen
  \bibfield  {author} {\bibinfo {author} {\bibfnamefont {R.}~\bibnamefont
  {Tomasello}}, \bibinfo {author} {\bibfnamefont {S.}~\bibnamefont {Komineas}},
  \bibinfo {author} {\bibfnamefont {G.}~\bibnamefont {Siracusano}}, \bibinfo
  {author} {\bibfnamefont {M.}~\bibnamefont {Carpentieri}},\ and\ \bibinfo
  {author} {\bibfnamefont {G.}~\bibnamefont {Finocchio}},\ }\href@noop {}
  {\bibfield  {journal} {\bibinfo  {journal} {Physical Review B}\ }\textbf
  {\bibinfo {volume} {98}},\ \bibinfo {pages} {024421} (\bibinfo {year}
  {2018})}\BibitemShut {NoStop}%
\bibitem [{\citenamefont {Wang}\ \emph {et~al.}(2018)\citenamefont {Wang},
  \citenamefont {Gan}, \citenamefont {Martinez}, \citenamefont {Tan},
  \citenamefont {Jalil},\ and\ \citenamefont {Lew}}]{R30}%
  \BibitemOpen
  \bibfield  {author} {\bibinfo {author} {\bibfnamefont {X.}~\bibnamefont
  {Wang}}, \bibinfo {author} {\bibfnamefont {W.}~\bibnamefont {Gan}}, \bibinfo
  {author} {\bibfnamefont {J.}~\bibnamefont {Martinez}}, \bibinfo {author}
  {\bibfnamefont {F.}~\bibnamefont {Tan}}, \bibinfo {author} {\bibfnamefont
  {M.}~\bibnamefont {Jalil}},\ and\ \bibinfo {author} {\bibfnamefont
  {W.}~\bibnamefont {Lew}},\ }\href@noop {} {\bibfield  {journal} {\bibinfo
  {journal} {Nanoscale}\ }\textbf {\bibinfo {volume} {10}},\ \bibinfo {pages}
  {733} (\bibinfo {year} {2018})}\BibitemShut {NoStop}%
\bibitem [{\citenamefont {Ang}\ \emph {et~al.}(2019)\citenamefont {Ang},
  \citenamefont {Gan},\ and\ \citenamefont {Lew}}]{R31}%
  \BibitemOpen
  \bibfield  {author} {\bibinfo {author} {\bibfnamefont {C.~C.~I.}\
  \bibnamefont {Ang}}, \bibinfo {author} {\bibfnamefont {W.}~\bibnamefont
  {Gan}},\ and\ \bibinfo {author} {\bibfnamefont {W.~S.}\ \bibnamefont {Lew}},\
  }\href@noop {} {\bibfield  {journal} {\bibinfo  {journal} {New Journal of
  Physics}\ }\textbf {\bibinfo {volume} {21}},\ \bibinfo {pages} {043006}
  (\bibinfo {year} {2019})}\BibitemShut {NoStop}%
\bibitem [{\citenamefont {Ma}\ \emph {et~al.}(2018)\citenamefont {Ma},
  \citenamefont {Zhang}, \citenamefont {Xia}, \citenamefont {Ezawa},
  \citenamefont {Jiang}, \citenamefont {Ono}, \citenamefont {Piramanayagam},
  \citenamefont {Morisako}, \citenamefont {Zhou},\ and\ \citenamefont
  {Liu}}]{R32}%
  \BibitemOpen
  \bibfield  {author} {\bibinfo {author} {\bibfnamefont {C.}~\bibnamefont
  {Ma}}, \bibinfo {author} {\bibfnamefont {X.}~\bibnamefont {Zhang}}, \bibinfo
  {author} {\bibfnamefont {J.}~\bibnamefont {Xia}}, \bibinfo {author}
  {\bibfnamefont {M.}~\bibnamefont {Ezawa}}, \bibinfo {author} {\bibfnamefont
  {W.}~\bibnamefont {Jiang}}, \bibinfo {author} {\bibfnamefont
  {T.}~\bibnamefont {Ono}}, \bibinfo {author} {\bibfnamefont {S.}~\bibnamefont
  {Piramanayagam}}, \bibinfo {author} {\bibfnamefont {A.}~\bibnamefont
  {Morisako}}, \bibinfo {author} {\bibfnamefont {Y.}~\bibnamefont {Zhou}},\
  and\ \bibinfo {author} {\bibfnamefont {X.}~\bibnamefont {Liu}},\ }\href@noop
  {} {\bibfield  {journal} {\bibinfo  {journal} {Nano letters}\ }\textbf
  {\bibinfo {volume} {19}},\ \bibinfo {pages} {353} (\bibinfo {year}
  {2018})}\BibitemShut {NoStop}%
\bibitem [{\citenamefont {Zhou}\ \emph {et~al.}(2019)\citenamefont {Zhou},
  \citenamefont {Mansell},\ and\ \citenamefont {van Dijken}}]{R33}%
  \BibitemOpen
  \bibfield  {author} {\bibinfo {author} {\bibfnamefont {Y.}~\bibnamefont
  {Zhou}}, \bibinfo {author} {\bibfnamefont {R.}~\bibnamefont {Mansell}},\ and\
  \bibinfo {author} {\bibfnamefont {S.}~\bibnamefont {van Dijken}},\
  }\href@noop {} {\bibfield  {journal} {\bibinfo  {journal} {Scientific
  Reports}\ }\textbf {\bibinfo {volume} {9}},\ \bibinfo {pages} {1} (\bibinfo
  {year} {2019})}\BibitemShut {NoStop}%
\bibitem [{\citenamefont {Qiu}\ \emph {et~al.}(2020)\citenamefont {Qiu},
  \citenamefont {Xia}, \citenamefont {Feng}, \citenamefont {Shen},
  \citenamefont {Morvan}, \citenamefont {Zhang}, \citenamefont {Liu},
  \citenamefont {Xie}, \citenamefont {Zhou},\ and\ \citenamefont {Zhao}}]{R34}%
  \BibitemOpen
  \bibfield  {author} {\bibinfo {author} {\bibfnamefont {L.}~\bibnamefont
  {Qiu}}, \bibinfo {author} {\bibfnamefont {J.}~\bibnamefont {Xia}}, \bibinfo
  {author} {\bibfnamefont {Y.}~\bibnamefont {Feng}}, \bibinfo {author}
  {\bibfnamefont {L.}~\bibnamefont {Shen}}, \bibinfo {author} {\bibfnamefont
  {F.~J.}\ \bibnamefont {Morvan}}, \bibinfo {author} {\bibfnamefont
  {X.}~\bibnamefont {Zhang}}, \bibinfo {author} {\bibfnamefont
  {X.}~\bibnamefont {Liu}}, \bibinfo {author} {\bibfnamefont {L.}~\bibnamefont
  {Xie}}, \bibinfo {author} {\bibfnamefont {Y.}~\bibnamefont {Zhou}},\ and\
  \bibinfo {author} {\bibfnamefont {G.}~\bibnamefont {Zhao}},\ }\href@noop {}
  {\bibfield  {journal} {\bibinfo  {journal} {Journal of Magnetism and Magnetic
  Materials}\ }\textbf {\bibinfo {volume} {496}},\ \bibinfo {pages} {165922}
  (\bibinfo {year} {2020})}\BibitemShut {NoStop}%
\bibitem [{\citenamefont {Li}\ \emph {et~al.}(2020)\citenamefont {Li},
  \citenamefont {Jin}, \citenamefont {Wen}, \citenamefont {Zhang},
  \citenamefont {Qin},\ and\ \citenamefont {Liu}}]{R35}%
  \BibitemOpen
  \bibfield  {author} {\bibinfo {author} {\bibfnamefont {W.}~\bibnamefont
  {Li}}, \bibinfo {author} {\bibfnamefont {Z.}~\bibnamefont {Jin}}, \bibinfo
  {author} {\bibfnamefont {D.}~\bibnamefont {Wen}}, \bibinfo {author}
  {\bibfnamefont {X.}~\bibnamefont {Zhang}}, \bibinfo {author} {\bibfnamefont
  {M.}~\bibnamefont {Qin}},\ and\ \bibinfo {author} {\bibfnamefont {J.-M.}\
  \bibnamefont {Liu}},\ }\href@noop {} {\bibfield  {journal} {\bibinfo
  {journal} {Physical Review B}\ }\textbf {\bibinfo {volume} {101}},\ \bibinfo
  {pages} {024414} (\bibinfo {year} {2020})}\BibitemShut {NoStop}%
\bibitem [{\citenamefont {Rybakov}\ and\ \citenamefont {Kiselev}(2019)}]{R36}%
  \BibitemOpen
  \bibfield  {author} {\bibinfo {author} {\bibfnamefont {F.~N.}\ \bibnamefont
  {Rybakov}}\ and\ \bibinfo {author} {\bibfnamefont {N.~S.}\ \bibnamefont
  {Kiselev}},\ }\href@noop {} {\bibfield  {journal} {\bibinfo  {journal}
  {Physical Review B}\ }\textbf {\bibinfo {volume} {99}},\ \bibinfo {pages}
  {064437} (\bibinfo {year} {2019})}\BibitemShut {NoStop}%
\bibitem [{\citenamefont {Foster}\ \emph {et~al.}(2019)\citenamefont {Foster},
  \citenamefont {Kind}, \citenamefont {Ackerman}, \citenamefont {Tai},
  \citenamefont {Dennis},\ and\ \citenamefont {Smalyukh}}]{R37}%
  \BibitemOpen
  \bibfield  {author} {\bibinfo {author} {\bibfnamefont {D.}~\bibnamefont
  {Foster}}, \bibinfo {author} {\bibfnamefont {C.}~\bibnamefont {Kind}},
  \bibinfo {author} {\bibfnamefont {P.~J.}\ \bibnamefont {Ackerman}}, \bibinfo
  {author} {\bibfnamefont {J.-S.~B.}\ \bibnamefont {Tai}}, \bibinfo {author}
  {\bibfnamefont {M.~R.}\ \bibnamefont {Dennis}},\ and\ \bibinfo {author}
  {\bibfnamefont {I.~I.}\ \bibnamefont {Smalyukh}},\ }\href@noop {} {\bibfield
  {journal} {\bibinfo  {journal} {Nature Physics}\ }\textbf {\bibinfo {volume}
  {15}},\ \bibinfo {pages} {655} (\bibinfo {year} {2019})}\BibitemShut
  {NoStop}%
\bibitem [{\citenamefont {Kind}\ \emph {et~al.}(2020)\citenamefont {Kind},
  \citenamefont {Friedemann},\ and\ \citenamefont {Read}}]{R38}%
  \BibitemOpen
  \bibfield  {author} {\bibinfo {author} {\bibfnamefont {C.}~\bibnamefont
  {Kind}}, \bibinfo {author} {\bibfnamefont {S.}~\bibnamefont {Friedemann}},\
  and\ \bibinfo {author} {\bibfnamefont {D.}~\bibnamefont {Read}},\ }\href@noop
  {} {\bibfield  {journal} {\bibinfo  {journal} {Applied Physics Letters}\
  }\textbf {\bibinfo {volume} {116}},\ \bibinfo {pages} {022413} (\bibinfo
  {year} {2020})}\BibitemShut {NoStop}%
\bibitem [{\citenamefont {Zeng}\ \emph {et~al.}(2020)\citenamefont {Zeng},
  \citenamefont {Zhang}, \citenamefont {Jin}, \citenamefont {Wang},
  \citenamefont {Song}, \citenamefont {Ma}, \citenamefont {Liu},\ and\
  \citenamefont {Wang}}]{R39}%
  \BibitemOpen
  \bibfield  {author} {\bibinfo {author} {\bibfnamefont {Z.}~\bibnamefont
  {Zeng}}, \bibinfo {author} {\bibfnamefont {C.}~\bibnamefont {Zhang}},
  \bibinfo {author} {\bibfnamefont {C.}~\bibnamefont {Jin}}, \bibinfo {author}
  {\bibfnamefont {J.}~\bibnamefont {Wang}}, \bibinfo {author} {\bibfnamefont
  {C.}~\bibnamefont {Song}}, \bibinfo {author} {\bibfnamefont {Y.}~\bibnamefont
  {Ma}}, \bibinfo {author} {\bibfnamefont {Q.}~\bibnamefont {Liu}},\ and\
  \bibinfo {author} {\bibfnamefont {J.}~\bibnamefont {Wang}},\ }\href@noop {}
  {\bibfield  {journal} {\bibinfo  {journal} {Applied Physics Letters}\
  }\textbf {\bibinfo {volume} {117}},\ \bibinfo {pages} {172404} (\bibinfo
  {year} {2020})}\BibitemShut {NoStop}%
\bibitem [{\citenamefont {Kind}\ and\ \citenamefont {Foster}(2021)}]{R40}%
  \BibitemOpen
  \bibfield  {author} {\bibinfo {author} {\bibfnamefont {C.}~\bibnamefont
  {Kind}}\ and\ \bibinfo {author} {\bibfnamefont {D.}~\bibnamefont {Foster}},\
  }\href@noop {} {\bibfield  {journal} {\bibinfo  {journal} {Physical Review
  B}\ }\textbf {\bibinfo {volume} {103}},\ \bibinfo {pages} {L100413} (\bibinfo
  {year} {2021})}\BibitemShut {NoStop}%
\bibitem [{\citenamefont {Tang}\ \emph {et~al.}(2021)\citenamefont {Tang},
  \citenamefont {Wu}, \citenamefont {Wang}, \citenamefont {Kong}, \citenamefont
  {Lv}, \citenamefont {Wei}, \citenamefont {Zang}, \citenamefont {Tian},\ and\
  \citenamefont {Du}}]{R41}%
  \BibitemOpen
  \bibfield  {author} {\bibinfo {author} {\bibfnamefont {J.}~\bibnamefont
  {Tang}}, \bibinfo {author} {\bibfnamefont {Y.}~\bibnamefont {Wu}}, \bibinfo
  {author} {\bibfnamefont {W.}~\bibnamefont {Wang}}, \bibinfo {author}
  {\bibfnamefont {L.}~\bibnamefont {Kong}}, \bibinfo {author} {\bibfnamefont
  {B.}~\bibnamefont {Lv}}, \bibinfo {author} {\bibfnamefont {W.}~\bibnamefont
  {Wei}}, \bibinfo {author} {\bibfnamefont {J.}~\bibnamefont {Zang}}, \bibinfo
  {author} {\bibfnamefont {M.}~\bibnamefont {Tian}},\ and\ \bibinfo {author}
  {\bibfnamefont {H.}~\bibnamefont {Du}},\ }\href@noop {} {\bibfield  {journal}
  {\bibinfo  {journal} {Nature Nanotechnology}\ }\textbf {\bibinfo {volume}
  {16}},\ \bibinfo {pages} {1086} (\bibinfo {year} {2021})}\BibitemShut
  {NoStop}%
\bibitem [{\citenamefont {Chen}\ \emph {et~al.}(2020)\citenamefont {Chen},
  \citenamefont {Li}, \citenamefont {Pavlidis},\ and\ \citenamefont
  {Moutafis}}]{RN1}%
  \BibitemOpen
  \bibfield  {author} {\bibinfo {author} {\bibfnamefont {R.}~\bibnamefont
  {Chen}}, \bibinfo {author} {\bibfnamefont {Y.}~\bibnamefont {Li}}, \bibinfo
  {author} {\bibfnamefont {V.~F.}\ \bibnamefont {Pavlidis}},\ and\ \bibinfo
  {author} {\bibfnamefont {C.}~\bibnamefont {Moutafis}},\ }\href@noop {}
  {\bibfield  {journal} {\bibinfo  {journal} {Physical Review Research}\
  }\textbf {\bibinfo {volume} {2}},\ \bibinfo {pages} {043312} (\bibinfo {year}
  {2020})}\BibitemShut {NoStop}%
\bibitem [{\citenamefont {Chen}\ \emph {et~al.}(2022)\citenamefont {Chen},
  \citenamefont {Li}, \citenamefont {Pavlidis},\ and\ \citenamefont
  {Moutafis}}]{RN2}%
  \BibitemOpen
  \bibfield  {author} {\bibinfo {author} {\bibfnamefont {R.}~\bibnamefont
  {Chen}}, \bibinfo {author} {\bibfnamefont {Y.}~\bibnamefont {Li}}, \bibinfo
  {author} {\bibfnamefont {V.~F.}\ \bibnamefont {Pavlidis}},\ and\ \bibinfo
  {author} {\bibfnamefont {C.}~\bibnamefont {Moutafis}},\ }\href@noop {}
  {\bibfield  {journal} {\bibinfo  {journal} {arXiv preprint arXiv:2203.13711}\
  } (\bibinfo {year} {2022})}\BibitemShut {NoStop}%
\bibitem [{\citenamefont {Kuchkin}\ \emph {et~al.}(2021)\citenamefont
  {Kuchkin}, \citenamefont {Chichay}, \citenamefont {Barton-Singer},
  \citenamefont {Rybakov}, \citenamefont {Bl{\"u}gel}, \citenamefont
  {Schroers},\ and\ \citenamefont {Kiselev}}]{R43}%
  \BibitemOpen
  \bibfield  {author} {\bibinfo {author} {\bibfnamefont {V.~M.}\ \bibnamefont
  {Kuchkin}}, \bibinfo {author} {\bibfnamefont {K.}~\bibnamefont {Chichay}},
  \bibinfo {author} {\bibfnamefont {B.}~\bibnamefont {Barton-Singer}}, \bibinfo
  {author} {\bibfnamefont {F.~N.}\ \bibnamefont {Rybakov}}, \bibinfo {author}
  {\bibfnamefont {S.}~\bibnamefont {Bl{\"u}gel}}, \bibinfo {author}
  {\bibfnamefont {B.~J.}\ \bibnamefont {Schroers}},\ and\ \bibinfo {author}
  {\bibfnamefont {N.~S.}\ \bibnamefont {Kiselev}},\ }\href@noop {} {\bibfield
  {journal} {\bibinfo  {journal} {Physical Review B}\ }\textbf {\bibinfo
  {volume} {104}},\ \bibinfo {pages} {165116} (\bibinfo {year}
  {2021})}\BibitemShut {NoStop}%
\bibitem [{\citenamefont {Vansteenkiste}\ \emph {et~al.}(2014)\citenamefont
  {Vansteenkiste}, \citenamefont {Leliaert}, \citenamefont {Dvornik},
  \citenamefont {Helsen}, \citenamefont {Garcia-Sanchez},\ and\ \citenamefont
  {Van~Waeyenberge}}]{R44}%
  \BibitemOpen
  \bibfield  {author} {\bibinfo {author} {\bibfnamefont {A.}~\bibnamefont
  {Vansteenkiste}}, \bibinfo {author} {\bibfnamefont {J.}~\bibnamefont
  {Leliaert}}, \bibinfo {author} {\bibfnamefont {M.}~\bibnamefont {Dvornik}},
  \bibinfo {author} {\bibfnamefont {M.}~\bibnamefont {Helsen}}, \bibinfo
  {author} {\bibfnamefont {F.}~\bibnamefont {Garcia-Sanchez}},\ and\ \bibinfo
  {author} {\bibfnamefont {B.}~\bibnamefont {Van~Waeyenberge}},\ }\href@noop {}
  {\bibfield  {journal} {\bibinfo  {journal} {AIP advances}\ }\textbf {\bibinfo
  {volume} {4}},\ \bibinfo {pages} {107133} (\bibinfo {year}
  {2014})}\BibitemShut {NoStop}%
\bibitem [{\citenamefont {Zeng}\ \emph {et~al.}(2022)\citenamefont {Zeng},
  \citenamefont {Song}, \citenamefont {Wang},\ and\ \citenamefont {Liu}}]{R42}%
  \BibitemOpen
  \bibfield  {author} {\bibinfo {author} {\bibfnamefont {Z.}~\bibnamefont
  {Zeng}}, \bibinfo {author} {\bibfnamefont {C.}~\bibnamefont {Song}}, \bibinfo
  {author} {\bibfnamefont {J.}~\bibnamefont {Wang}},\ and\ \bibinfo {author}
  {\bibfnamefont {Q.}~\bibnamefont {Liu}},\ }\href@noop {} {\bibfield
  {journal} {\bibinfo  {journal} {Journal of Physics D: Applied Physics}\
  }\textbf {\bibinfo {volume} {55}},\ \bibinfo {pages} {185001} (\bibinfo
  {year} {2022})}\BibitemShut {NoStop}%
\bibitem [{\citenamefont {Thiele}(1973)}]{R45}%
  \BibitemOpen
  \bibfield  {author} {\bibinfo {author} {\bibfnamefont {A.}~\bibnamefont
  {Thiele}},\ }\href@noop {} {\bibfield  {journal} {\bibinfo  {journal}
  {Physical Review Letters}\ }\textbf {\bibinfo {volume} {30}},\ \bibinfo
  {pages} {230} (\bibinfo {year} {1973})}\BibitemShut {NoStop}%
\bibitem [{\citenamefont {Kolesnikov}\ \emph {et~al.}(2018)\citenamefont
  {Kolesnikov}, \citenamefont {Stebliy}, \citenamefont {Samardak},\ and\
  \citenamefont {Ognev}}]{R46}%
  \BibitemOpen
  \bibfield  {author} {\bibinfo {author} {\bibfnamefont {A.~G.}\ \bibnamefont
  {Kolesnikov}}, \bibinfo {author} {\bibfnamefont {M.~E.}\ \bibnamefont
  {Stebliy}}, \bibinfo {author} {\bibfnamefont {A.~S.}\ \bibnamefont
  {Samardak}},\ and\ \bibinfo {author} {\bibfnamefont {A.~V.}\ \bibnamefont
  {Ognev}},\ }\href@noop {} {\bibfield  {journal} {\bibinfo  {journal}
  {Scientific reports}\ }\textbf {\bibinfo {volume} {8}},\ \bibinfo {pages} {1}
  (\bibinfo {year} {2018})}\BibitemShut {NoStop}%
\bibitem [{\citenamefont {Lai}\ \emph {et~al.}(2017)\citenamefont {Lai},
  \citenamefont {Zhao}, \citenamefont {Tang}, \citenamefont {Ran},
  \citenamefont {Wu}, \citenamefont {Xia}, \citenamefont {Zhang},\ and\
  \citenamefont {Zhou}}]{R47}%
  \BibitemOpen
  \bibfield  {author} {\bibinfo {author} {\bibfnamefont {P.}~\bibnamefont
  {Lai}}, \bibinfo {author} {\bibfnamefont {G.}~\bibnamefont {Zhao}}, \bibinfo
  {author} {\bibfnamefont {H.}~\bibnamefont {Tang}}, \bibinfo {author}
  {\bibfnamefont {N.}~\bibnamefont {Ran}}, \bibinfo {author} {\bibfnamefont
  {S.}~\bibnamefont {Wu}}, \bibinfo {author} {\bibfnamefont {J.}~\bibnamefont
  {Xia}}, \bibinfo {author} {\bibfnamefont {X.}~\bibnamefont {Zhang}},\ and\
  \bibinfo {author} {\bibfnamefont {Y.}~\bibnamefont {Zhou}},\ }\href@noop {}
  {\bibfield  {journal} {\bibinfo  {journal} {Scientific reports}\ }\textbf
  {\bibinfo {volume} {7}},\ \bibinfo {pages} {1} (\bibinfo {year}
  {2017})}\BibitemShut {NoStop}%
\end{thebibliography}%

\end{document}